\documentclass[prl,aps,twocolumn,showpacs,floatfix]{revtex4}
\usepackage{graphicx}
\usepackage{amsmath}

\begin{document}

\title{Susceptibility of the 2D $S=\frac{1}{2}$ Heisenberg antiferromagnet 
with an impurity}

\author{Kaj H. H\"oglund} 
\author{Anders W. Sandvik} 
\affiliation{Department of Physics, {\AA}bo Akademi University, 
Porthansgatan 3, FIN-20500, Turku, Finland}

\date{\today}

\pacs{75.10.Jm, 75.10.Nr, 75.40.Cx, 75.40.Mg}

\begin{abstract}
We use a quantum Monte Carlo method (stochastic series expansion) to study 
the effects of a magnetic or nonmagnetic impurity on the magnetic 
susceptibility of the two-dimensional Heisenberg antiferromagnet. At low
temperatures, we find a log-divergent contribution to the transverse 
susceptibility. We also introduce an effective few-spin model that can 
quantitatively capture the differences between magnetic and nonmagnetic
impurities at high and intermediate temperatures.
\end{abstract}

\maketitle

Static impurities can be introduced into the CuO${\rm _2}$ planes of the 
high-$T_{c}$ cuprates as a means of probing their electronic correlations and 
excitations; some unwanted impurities and defects are also always present in 
``pure'' systems. Impurities coupled to a two-dimensional (2D) host system 
hence constitute an important class of quantum many-body problems. When Cu 
ions in hole-doped CuO$_{\rm 2}$ planes are substituted with Zn, experiments 
indicate that local magnetic moments are induced at the Cu sites neighboring 
the nonmagnetic Zn impurities \cite{mahajan}. This behavior can be reproduced 
in spin-gapped Heisenberg antiferromagnets \cite{ng}, e.g., in systems of 
weakly coupled spin-$\frac{1}{2}$ ladders \cite{sandvik,martins,sachdev}. 
Although spin models cannot address questions specifically related to a 
metallic or superconducting host, static impurities in various gapped 
(paramagnetic) and gapless (antiferromagnetic) Heisenberg systems are 
important limiting models for understanding the physics of quantum spin 
defects. They also have direct experimental realizations in the cuprates
\cite{doping,greven}. 

Several studies have addressed static vacancies 
\cite{ng,sandvik,sachdev,bulut,nagaosa} and added spins \cite{igarashi,kotov}
in 2D $S=1/2$ Heisenberg models. The localized moments forming in 
spin-gapped systems are now well understood. Recently, a universal 
behavior was predicted \cite{sachdev} for the $T>0$ impurity effects in 
systems that have long-range order at $T=0$ (i.e., in the ``renormalized 
classical'' regime \cite{chn}), as well as in systems close to a 
quantum-critical point. Some predictions for quantum-critical systems 
have been confirmed numerically \cite{sachdev2}, whereas other aspects of 
the theory remain unsettled \cite{troyer,sushkov}. The predictions in the 
renormalized-classical regime have not yet been tested.

In this Letter, we consider the 2D square-lattice spin-$\frac{1}{2}$ Heisenberg
antiferromagnet with \textit{(i)} a vacancy (nonmagnetic impurity) or 
\textit{(ii)} an off-plane added spin (magnetic impurity) coupled to a 
single host spin. We perform quantum Monte Carlo 
calculations, using the stochastic series expansion (SSE) technique 
\cite{sse}, and determine the effects of the two different impurities on the 
uniform magnetic susceptibility. Our results confirm quantitatively a 
classical-like longitudinal Curie contribution resulting from alignment of the
impurity moment with the local N\'eel order \cite{sachdev}. However, we also 
find evidence of a logarithmic divergence as $T\to 0$ in the 
transverse component of the impurity susceptibility, instead of the predicted 
$T$ independence at low $T$. We point out qualitative differences between 
the vacancy and the added spin at high and intermediate $T$ and explain these 
in terms of an effective few-spin model.

We begin by defining three spin-$\frac{1}{2}$  Hamiltonians describing the 
models investigated:
\begin{subequations}
\begin{eqnarray}
H_{0}&=&J\sum_{\langle i,j\rangle}\mathbf{S}_{i}\cdot \mathbf{S}_{j},
	\label{eq:clean}\\
H_{-}&=&J\sum_{\genfrac{}{}{0pt}{}{\langle i,j\rangle}{i,j\not=0}}
         \mathbf{S}_{i}\cdot \mathbf{S}_{j},
	 \label{eq:vacancy}\\
H_{+}&=&J\sum_{\langle i,j\rangle}\mathbf{S}_{i}\cdot \mathbf{S}_{j}
	+J_{\perp}\mathbf{S}_{a}\cdot \mathbf{S}_{0}, \label{eq:added}
\end{eqnarray}		
\label{fullmodels}
\end{subequations}
where $\langle i,j\rangle$ denotes a pair of nearest-neighbor sites on a 
periodic $L\times L$ lattice. $H_{0}$ is the standard Heisenberg model.
In $H_{-}$ the spin $\mathbf{S}_0$ has been removed, creating a vacancy. We 
also study systems with two vacancies at maximum separation on different 
sublattices. In $H_{+}$ an added spin-$\frac{1}{2}$, $\mathbf{S}_{a}$, is 
coupled to one of the spins, $\mathbf{S}_{0}$, in the plane. We here only 
consider $J_{\perp}=J$. 

We have calculated the total susceptibilities
\begin{equation}\label{eq:ums}
\chi ^{z}_{k} =\frac{1}{T} \left \langle \left 
(\sum_{i=1}^N S_{i}^{z}\right )^{2}\right \rangle,
\end{equation}
with $k=0,-,+$ corresponding to $H_0$, $H_-$, and $H_+$. We have used 
square lattices with $L=4,8,16,32$, and $64$, at temperatures down to 
$T=J/32$. The  number of spins $N=L^2$ for $k=0$ and $L^2 \pm 1$ for $k=\pm$. 
In order to determine the effects of the impurities, we follow 
Ref.~\onlinecite{sachdev} and define the impurity susceptibilities
\begin{equation}
\chi _{\text{imp}}^{z,\pm}=\chi ^{z}_{\pm}-\chi^{z}_{0}.
\label{ximp}
\end{equation}
In the case of two vacancies, the impurity susceptibility is further
normalized by a factor $\frac{1}{2}$, so that for $L\to \infty$ it
should be the same as for a single vacancy.

The temperature dependence of $\chi _{\text{imp}}^{z}$ was discussed by
Sachdev {\it et al.} \cite{sachdev} on the basis of quantum field theory.
They concluded that at $T=0$, the component $\chi _{\text{imp}}^{\parallel}$
parallel to the  direction of the N\'eel order vanishes, while the
perpendicular component $\chi _{\text{imp}} ^{\perp}=C_{3}/\rho _{s}$,
where $C_{3}$ is a constant and $\rho _{s}$ is the spin stiffness
of the host. For $T>0$ the correlation length $\xi$ grows exponentially as
$T\to 0$ \cite{chn}. For a general impurity spin, a moment of magnitude
$S$ was then predicted to result from the alignment of ${\bf S}$ with a large 
N\'eel-ordered domain, i.e., it does not have the quantum mechanical 
magnitude $\sqrt{S(S+1)}$ which might have been naively expected. The full 
low-$T$ impurity susceptibility is hence $\chi _{\text{imp}}^{z}=\frac{1}{3}
\frac{1}{T}S^{2}+\frac{2}{3} \chi _{\text{imp}}^{\perp}$ \cite{sachdev}. 

In order to have a simple system which reproduces this expected low-$T$ 
behavior, and also accounts for non-universal behavior at higher $T$ for 
different types of impurities, we introduce a simple effective few-spin
model. The idea is to model a large N\'eel-ordered region by a classical 
vector $\mathbf{N}$. Three Hamiltonians corresponding to the original 
models (\ref{fullmodels}) are defined:
\begin{subequations}
\begin{eqnarray}
H_{0}^{\rm eff}&=&\alpha \mathbf{S}_{0}\cdot \mathbf{S}_{e}+r\mathbf{N}\cdot 
	\mathbf{S}_{e}-h_{z}M_{0}^{z}, \label{eq:undoped}\\
H_{-}^{\rm eff}&=&r\mathbf{N}\cdot \mathbf{S}_{e}-h_{z}M_{-}^{z}, 
	\label{eq:nonmagnetic}\\
H_{+}^{\rm eff}&=&\alpha \mathbf{S}_{0}\cdot \mathbf{S}_{e}+
r\mathbf{N}\cdot \mathbf{S}_{e} + J_{\perp}\mathbf{S}_{a}\cdot \mathbf{S}_{0}
-h_{z}M_{+}^{z}.~~~ \label{eq:magnetic}
\end{eqnarray}		
\label{effmodels}
\end{subequations}
The effective Hamiltonian for the vacancy model, $H_{-}^{\rm eff}$, contains
a single spin $\mathbf{S}_{e}$ representing a remnant spin-$\frac{1}{2}$ due
to the sublattice asymmetry caused by the vacancy. It is coupled to the
unit vector $\mathbf{N}$, representing the orientation of the local 
N\'eel order of the host antiferromagnet. The magnitude of this order is 
absorbed into the effective coupling $r$. In $H_{0}^{\rm eff}$ we 
``reinsert'' the spin $\mathbf{S}_{0}$ that was removed in the vacancy 
model, and couple it to $\mathbf{S}_{e}$. Further, including the Heisenberg 
interaction between the host spin $\mathbf{S}_{0}$ and the added spin 
$\mathbf{S}_{a}$, we arrive at the effective Hamiltonian for the added 
spin model $H_{+}^{\rm eff}$.

We determine the impurity susceptibilities (\ref{ximp}) of the effective 
models in the same way as for the original models. In Eqs.~(\ref{effmodels}) 
we have explicitly indicated how an applied external field, which defines 
the $z$ direction, couples to the systems. The magnetization operators 
are  $M_{-}^{z}=S_{e}^{z}$, $M_{0}^{z}=
S_{0}^{z}+S_{e}^{z}$, and $M_{+}^{z}=S_{a}^{z}+ S_{0}^{z}+S_{e}^{z}$. 
The susceptibilities for $k=0,-,+$, corresponding to $H_{0}^{\rm eff}$,
$H_{-}^{\rm eff}$, and $H_{+}^{\rm eff}$, can be written in the form
\begin{equation}
\chi ^{z}_{k}=\frac{\partial}{\partial h_{z}}\langle M_{k}^{z}\rangle
	=\frac{1}{3}\chi ^{\parallel}_{k}+\frac{2}{3}\chi ^{\perp}_{k},
\label{eq:usdef}
\end{equation}
where $\parallel$ and $\perp$ refer to the directions parallel and 
perpendicular to the vector $\mathbf{N}$. After straight-forward 
diagonalization of the effective Hamiltonians in the $\parallel$ basis, 
the two components can be evaluated using
\begin{subequations}
\begin{eqnarray}
\chi^{\parallel}_{k}&=&\frac{1}{T} \langle (M_{k}^{\parallel})^2 \rangle ,
        \label{eq:linres1} \\
\chi^{\perp}_{k}&=&\int_{0}^{1/T}d\tau \langle
	M_{k}^{\perp}(\tau)M_{k}^{\perp}(0)\rangle .
	\label{eq:linres2}
\end{eqnarray}
\end{subequations}
In (\ref{eq:usdef}) an average over all orientations of $\mathbf{N}$ relative 
to the fixed $z$ axis has been taken. The coupling of the impurity spin to 
$\mathbf{N}$, therefore, and in accordance with Ref.~\onlinecite{sachdev}, 
leads to a classical-like low-$T$ divergence; $\chi^{z,\pm}_{\text{imp}} 
\to \frac{1}{12}\frac{1}{T} +\frac{2}{3}\chi ^{\perp}_{\pm}$, where the 
second term is constant at low $T$ (instead of having the $r=0$ form 
$\frac{1}{6}\frac{1}{T}$).

We do not attempt to derive values for the couplings $\alpha$ and $r$ 
[however, $J_\perp=J$ as in (\ref{eq:added})]. Moreover, at least $r$ should
in principle have some $T$ dependence. All results presented will be for 
constant $\alpha/J=2.1$ and $r/J=1.75$, which give a reasonable over-all
agreement with the SSE calculations. We will show that although the effective
model is highly simplified, it captures some of the differences 
between the vacancy and the added spin.

Next, we present the results of the SSE calculations for the models 
(\ref{fullmodels}) and compare with the corresponding effective models. The 
impurity susceptibilities of interest, Eq.~(\ref{ximp}), are defined as 
differences between two extensive quantities. Although improved estimators 
\cite{evertz} were utilized in the SSE, the relative statistical noise grows
very rapidly with increasing $L$. We are therefore currently limited to 
$L \le 64$ and $T \ge J/32$.

\begin{figure}
\includegraphics[width=8cm]{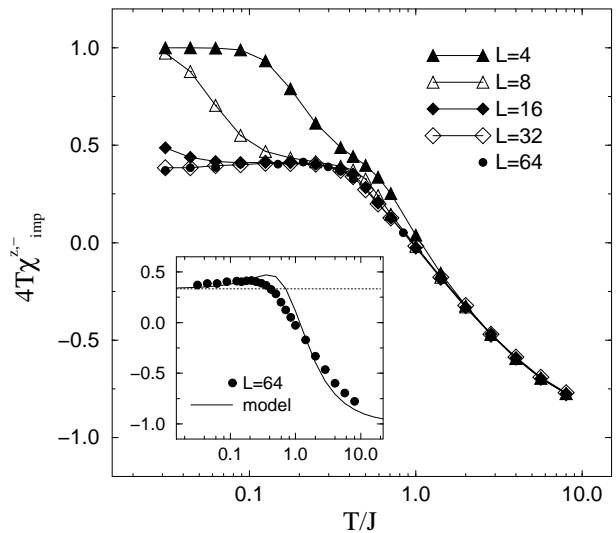}
\caption{Impurity susceptibilities for different system sizes $L$ with one 
vacancy. Error bars are smaller than the symbols. The inset shows a 
comparison between $L=64$ simulation data and the effective model.}
\label{d_bild}
\end{figure}

The single-vacancy impurity susceptibility multiplied by $4T$ is shown in 
Fig.~\ref{d_bild}. At high $T$ the data for different $L$ are 
indistinguishable, while at low $T$ finite-size effects are clearly seen 
for $L\le 16$. In the range of $T$ considered, all finite-size effects
are eliminated within statistical errors for $L=64$. The observed behavior 
at high $T$ is easily understood as the total susceptibility is then
just the sum of Curie contributions of each independent spin; 
$\chi _{\text{imp}}^{z,-}(T\to \infty)=(L^2-1)/4T-L^2/4T=-\frac{1}{4}
\frac{1}{T}$. At low $T$ the $S=\frac{1}{2}$ ground state of $H_-$ and $S=0$ 
of $H_0$ lead to the observed Curie behavior $\chi _{\text{imp}}^{z,-} = 
+\frac{1}{4}\frac{1}{T}$ for small $L$. As $L$ grows this 
finite-size effect vanishes, and we observe the predicted 
$\frac{1}{12}\frac{1}{T}$ contribution \cite{sachdev} arising 
from the longitudinal part, in analogy with the effective model. Hence the 
nearly constant $4T\chi _{\text{imp}}^{z,-} \approx \frac{1}{3}$ for $L=32$ 
and $64$. The effective model reproduces the behavior reasonably well, 
as shown in the inset of Fig.~\ref{d_bild}.

\begin{figure}
\includegraphics[width=8cm]{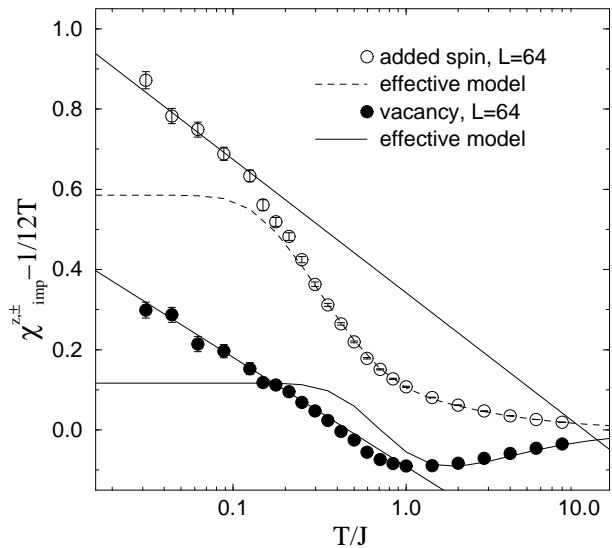}
\caption{$\chi_{\text{imp}}^{z,\pm}-\frac{1}{12}\frac{1}{T}$ for systems of 
size $L=64$ with a vacancy and an added spin. The straight lines
are fits to the low-$T$ simulation data.}
\label{new_bild}
\end{figure}

Looking more carefully at the SSE data, the predicted low-$T$ constant 
behavior in the quantity $\chi _{\text{imp}}^{z,-}-\frac{1}{12}\frac{1}{T}$, 
which should reduce to $\frac{2}{3}\chi _{\text{imp}}^{\perp ,-}$ as $T\to 0$,
is not observed. In Fig.~\ref{new_bild} (solid circles) we present 
results for $\chi _{\text{imp}}^{z,-}-\frac{1}{12}\frac{1}{T}$ for $L=64$
(data for smaller $L$ indicate that there are no significant finite-size 
effects for $L=64$). It shows an apparent logarithmically divergent behavior,
roughly from the onset $T$ of renormalized-classical behavior in the pure 
2D Heisenberg susceptibility, $T/J \approx 0.3$ \cite{kim}. This is also 
approximately where the corresponding effective-model result converges to 
a constant. 

\begin{figure}
\includegraphics[width=8cm]{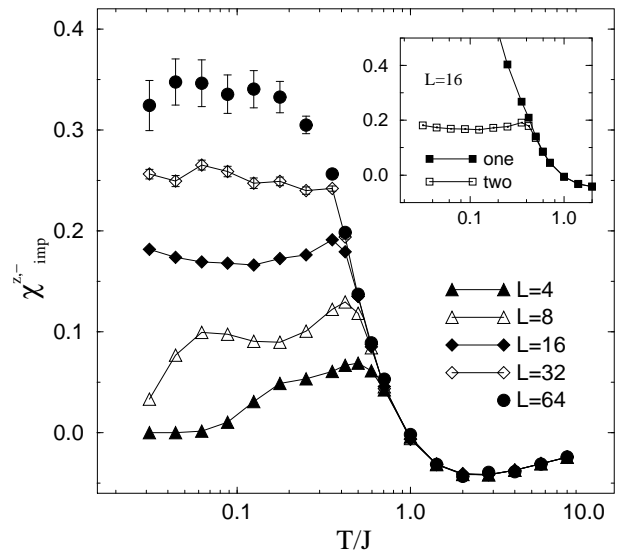}
\caption{Impurity susceptibility for systems with two vacancies. The inset 
shows results for $L=16$ systems with one and two vacancies.}
\label{d_2_bild}
\end{figure}

Results for two vacancies are shown in Fig.~\ref{d_2_bild}. The high-$T$ 
behavior has the same explanation as for the single vacancy. At low $T$, 
the moments due to the two vacancies, which reside on 
different sublattices, are pinned by the N\'eel order antiparallel to each 
other, resulting in a vanishing $\chi _{\text{imp}}^{\parallel ,-}$. Hence, 
in this case $\chi _{\text{imp}}^{z,-}$ does not diverge for small $L$ and 
we do not, therefore, multiply our results by $T$. The inset in 
Fig.~\ref{d_2_bild} shows a comparison between one and two vacancies for 
$L=16$. The $T$ at which the two curves deviate corresponds to a correlation 
length of the same order as the separation between the two impurities, 
$\xi \approx L/2$, i.e., above this $T$ the two impurities couple to 
different N\'eel domains and behave as independent vacancies. Since $\xi$ 
diverges exponentially, the point of deviation moves very slowly towards 
$T=0$ as $L$ grows. At low $T$ the resulting $\chi _{\text{imp}}^{z,-}$ for 
two vacancies shows little $T$ dependence for large $L$. However, no sign of 
convergence of the plateau value is seen as the system size grows. 
If $\chi_{\text{imp}}^{\perp ,-}$ is finite as $T\to 0$, we would 
expect such a convergence at low $T$ (with a peak at intermediate $T$ for 
very large $L$, where the cut-off of the $\frac{1}{12}\frac{1}{T}$ divergence
occurs at low $T$). It should be noted that in a finite system there will 
always be some interactions also between the $\perp$ components of the two 
impurity spins at low $T$. We can therefore not expect the behavior for two 
vacancies to be given in a straight-forward way by the single-vacancy 
results in Fig.~\ref{new_bild}. The roughly $\ln{(L)}$ divergence 
of the plateau height in Fig.~\ref{d_2_bild} is, however, fully in line 
with a log-divergent $\chi_{\text{imp}}^{\perp ,-}$ for a single vacancy. 

\begin{figure}
\includegraphics[width=8cm]{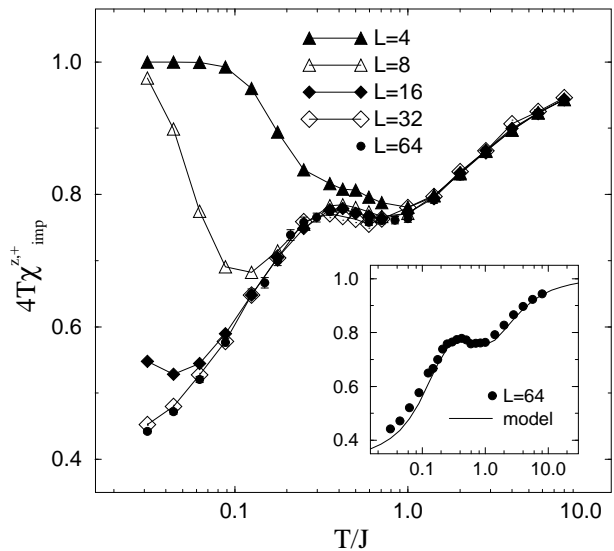}
\caption{Impurity susceptibilities for different system sizes $L$
with an off-plane added spin. The inset shows a comparison with
the effective model.}
\label{a_bild}
\end{figure}

Results for $\chi_{\text{imp}}^{z,+}$ for systems with an off-plane added 
spin are shown in Fig.~\ref{a_bild}. The high-$T$ behavior, as well as the 
low-$T$ behavior for small systems ($L=4,8$), is understood by the same 
arguments as for the vacancy. Note that in this case the high-$T$ impurity 
susceptibility is $+\frac{1}{4}\frac{1}{T}$, 
instead of $-\frac{1}{4}\frac{1}{T}$ for the 
vacancy. Again, we believe that all finite-size effects are eliminated 
for $L=64$ down to $T=J/32$. Contrary to the behavior for the vacancy  
in Fig.~\ref{d_bild}, the $T$ at which $4T\chi _{\text{imp}}^{z,+}$
assumes an almost
constant value $\frac{1}{3}$ has not yet been reached at $T=J/32$. Moreover, 
we note the substantial differences between $\chi _{\text{imp}}^{z,+}$ 
and $\chi_{\text{imp}}^{z,-}$ at intermediate $T$. In particular,
the shoulder seen in Fig.~\ref{a_bild} for $T \sim 0.2-1.2$ has no counterpart
in Fig.~\ref{d_bild}. This feature is clearly due to the internal structure of 
the added spin impurity and is reproduced very well by the effective model. 
In Fig.~\ref{new_bild} (open circles) we present the results for 
$\chi_{\text{imp}}^{z,+}-\frac{1}{12}\frac{1}{T}$ for $L=64$.
As in the vacancy 
case, it appears to be log-divergent at low $T$. Note also that the 
effective model describes the added spin impurity very well down to quite 
low temperatures. It captures the behavior better than for the vacancy, 
but a better agreement for the vacancy can also be achieved by using 
slightly different $\alpha$ and $r$.

Summarizing our results, we have confirmed the predicted \cite{sachdev}
contribution $\frac{1}{12}\frac{1}{T}$ to the impurity susceptibility. 
However, we also find a logarithmic divergence of 
$\chi _ {\text{imp}}^{z,\pm}-\frac{1}{12}\frac{1}{T}$, instead of the constant 
$\frac{2}{3}\chi _{\text{imp}}^{\perp}$ predicted for this quantity at low $T$ 
(and reproduced with our 
effective model). We cannot, of course, completely exclude an approach to a 
constant at still lower temperatures, but such a slow convergence had also not
been anticipated. We note that the separation of $\chi _{\text{imp}}^{z}$ into
transverse and longitudinal components, Eq.~(\ref{eq:usdef}), is strictly 
correct for the 2D Heisenberg model only at $T=0$. However, the very sudden 
cutoff of the divergence seen in the two-vacancy data in Fig.~\ref{d_2_bild}
supports the notion of a component aligning very strongly to the local N\'eel
order (which becomes the global order at the $L$-dependent crossover $T$) 
and justifies the relation $\frac{2}{3}\chi _{\text{imp}}^{\perp} = 
\chi _{\text{imp}}^{z}-\frac{1}{12}\frac{1}{T}$ also at relatively high 
temperatures. Hence, our results are most naturally interpreted as a 
log divergent $\chi _{\text{imp}}^{\perp}$. This is in fact in line with a 
Green's function calculation by Nagaosa \textit{et al.} \cite{nagaosa}. They 
found that for a system with a vacancy, the frequency dependent impurity 
susceptibility at $T=0$, $\chi_{\text{imp}} ^{\perp} (T=0,\omega)$, was 
log divergent when $\omega \to 0$. In view of the renormalized-classical 
\cite{chn} picture, this is consistent with exact results \cite{harris} for 
the classical 2D Heisenberg model. An anomalous perpendicular susceptibility
was also recently noted for the $S=1/2$ model at finite impurity 
concentration \cite{chernyshev}.

The results presented here call for a reexamination of the field theory 
\cite{sachdev} of quantum impurities in the renormalized-classical regime. 
Very recent efforts to explain the log divergence, motivated by our
numerical findings, have indicated that the impurity moment 
acquires a previously unnoticed correction of order $T \ln (1/T)$ to its 
leading order value $S$ in the renormalized-classical regime 
\cite{sachdev3,sushkov2}, and that this can be interpreted as a 
log-divergent contribution to $\chi_{\rm imp}^{\perp}$ as proposed here 
(results in the quantum-critical regime \cite{sachdev} remain unchanged). 

Apart from the log divergence, the effective few-spin model that we have 
introduced here gives a good description of the impurity susceptibility at 
high and intermediate $T$. In particular, it captures very well the
nonmonotonic $T$ dependence that we have found in the case of an added 
spin. 

We would like to thank Oleg Sushkov, and, in particular, Subir Sachdev for 
very useful correspondence and comments on the manuscript. This work is 
supported by the Academy of Finland, project No.~26175.

\end{document}